\documentclass[twocolumn,superscriptaddress,aps,prb,floatfix]{revtex4-1}

\usepackage{amsmath}
\usepackage{graphicx}
\usepackage[english]{babel}
\usepackage[version=4]{mhchem}
\usepackage{tabularx}
\usepackage{multirow}
\usepackage{color}
\usepackage{verbatim}
\usepackage{url}
\usepackage{xspace}
\usepackage{dcolumn}
\usepackage{multirow}
\usepackage{siunitx}

\newcolumntype{L}{>{\raggedright\arraybackslash}X}

\def\beq {\begin{equation}}
\def\eeq {\end{equation}}
\def\w {\omega}
\def\bfq {\mathbf{q}}

\def\bfG {\mathbf{G}}
\def\bfk {\mathbf{k}}
\def\bfr {\mathbf{r}}
\newcommand{\bra}[1]{\langle #1|}
\newcommand{\ket}[1]{|#1\rangle}
\def\cbn{$c$--BN\xspace}
\def\hbn{$h$--BN\xspace}
\def\rbn{$r$--BN\xspace}
\def\wbn{$w$--BN\xspace}

\newcommand{\lsi}{Laboratoire des Solides Irradi\'es, \'Ecole Polytechnique, CNRS, CEA,  Universit\'e Paris-Saclay, F-91128 Palaiseau, France}
\newcommand{\etsf}{European Theoretical Spectroscopy Facility (ETSF)}
\newcommand{\soleil}{Synchrotron SOLEIL, L'Orme des Merisiers, Saint-Aubin, BP 48, F-91192 Gif-sur-Yvette, France}

\begin{document}
\title{Cubic BN optical gap and intragap optically active defects}

\author{Anna Tararan}
\affiliation{Laboratoire de Physique des Solides, Univ. Paris-Sud, CNRS UMR
8502, F-91405, Orsay, France}

\author{Stefano di Sabatino}
\affiliation{\lsi}
\affiliation{\etsf}

\author{Matteo Gatti}
\affiliation{\lsi}
\affiliation{\etsf}
\affiliation{\soleil}

\author{Takashi Taniguchi}
\affiliation{National Institute for Materials Science, Namiki 1-1, Tsukuba, 
Ibaraki 305-0044, Japan}

\author{Kenji Watanabe}
\affiliation{National Institute for Materials Science, Namiki 1-1, Tsukuba, 
Ibaraki 305-0044, Japan}

\author{Lucia Reining}
\affiliation{\lsi}
\affiliation{\etsf}

\author{Luiz H. G. Tizei}
\affiliation{Laboratoire de Physique des Solides, Univ. Paris-Sud, CNRS UMR
8502, F-91405, Orsay, France}

\author{Mathieu Kociak}
\affiliation{Laboratoire de Physique des Solides, Univ. Paris-Sud, CNRS UMR
8502, F-91405, Orsay, France}

\author{Alberto Zobelli}
\email{alberto.zobelli@u-psud.fr}
\affiliation{Laboratoire de Physique des Solides, Univ. Paris-Sud, CNRS UMR 
8502, F-91405, Orsay, France}

\begin{abstract}
We report a comprehensive study on the optical properties of cubic boron nitride (\cbn) and its optically active defects. Using electron energy-loss spectroscopy (EELS) within a monochromated scanning transmission electron microscope (STEM) on the highest-quality  crystals available, we demonstrate unequivocally that the optical-gap energy of \cbn  slightly exceeds 10 eV. Further theoretical analysis in the framework of the Bethe-Salpeter equation of many-body perturbation theory supports this result. 
The spatial localization of defect-related emissions has been investigated using nanometric resolved cathodoluminescence (nano-CL) in a STEM. By high-temperature annealing a \cbn powder, we have promoted phase transitions in nanometric domains which have been detected by the appearance of specific hexagonal-phase signatures in both EELS and CL spectra. A high number of  intragap optically active centers are known in \cbn, but the literature is rather scattered and hence has been summarized here. For several emission lines
we have obtained nano-CL maps which show emission spot sizes as small as few tens of nanometers. Finally, by coupling nano-CL to a Hanbury-Brown-Twiss intensity interferometer, we have addressed individual spots in order to identify the possible presence of single-photon sources. The observed CL bunching effect  is compatible with a limited set of single-photon emitters and it permits obtaining emission lifetimes of the order of the nanosecond.
\end{abstract}

\date{\today}

\maketitle

\section{Introduction}

Group-III nitrides are wide band-gap semiconductors at the core of many modern optoelectronic applications. Currently, AlN, GaN, InN and ternary compounds are for instance fundamental for light-emitting devices. Due to the high stability of the \ce{N2} molecule, these materials are not found in nature. The synthesis of high-quality crystals has been  a challenge for a long time before these structures could be investigated for technological purposes. Boron nitride belongs to the same group of semiconductors but its interest as a potential optical material is still in its infancy. In the last twenty years, the synthesis of high-quality hexagonal BN (\hbn) crystals has permitted  important steps towards the comprehension of the peculiar optical properties of this wide band-gap layered material. However, studies on other boron-nitride phases are more incomplete due to the even more limited availability of high-quality crystals.

Cubic boron nitride (\cbn) is the zinc-blend allotropic form of boron nitride, also known as sphalerite phase. Boron and nitrogen are tetrahedrally coordinated forming a double-face-centered cubic lattice. The crystal structure appears thus isoelectronic with diamond, as \hbn is with graphite. The discovery of \cbn was inspired by the successful conversion of graphite into diamond, achieved at General Electrics in 1955.\cite{bundy_man-made_1955, hall_synthesis_1961} One year later, in the same research group, Robert H. Wentorf succeeded in the synthesis of \cbn (brand name Borazon) by heating boron and nitrogen mixtures at high pressure with a nitride catalyst.\cite{wentorf_cubic_1957, wentorf_preparation-cBN_1962} \cbn is now commonly synthesized by the temperature-gradient method at high pressure and high temperature (HPHT) in a variety of alkali or alkali-earth B--N solvents.\cite{mishima_electric_2000} Sintered \cbn is currently extensively used for cutting and abrasive tools, especially in presence of ferrous metals, which are able to degrade diamond.

Together with a clear interest due to its mechanical properties, the wide band gap of \cbn, by analogy with diamond and other wide-gap nitrides, makes it a very promising material for potential optical applications. 
\cbn shares interesting features with its hexagonal counterpart, such as an  intense intrinsic luminescence and bright and stable luminescent crystal defects. The quality of \cbn crystals is generally poor presenting a high density of point and extended defects, impurities or residual 
\hbn inclusions.\cite{taylor_observation_1994, matsumoto_introducing_2001, yu_morphology_2006}  In 2001, high-purity macroscopic crystals, with sizes of the order of 3 mm, were obtained using fresh barium boron nitride (\ce{Ba3B2N4}) as the solvent, with particular care to avoid solvent degradation by air and humidity.\cite{taniguchi_spontaneous_2001,
taniguchi_synthesis_2007} These high-quality samples can help in enlightening the intrinsic optical properties of \cbn and those of optically active defects.

The \cbn band gap is expected to exceed that of \hbn, but its precise value is still debated. Experimentally, a maximum estimation of 6.4 eV has been deduced from absorption and reflectance measurements.\cite{chrenko_ultraviolet_1974, mishima_electric_2000} Theoretically, the \cbn band structure has been found to display an indirect band gap. State-of-the-art quasiparticle (QP) calculations in the GW approximation\cite{Hedin1965} (GWA) have determined  a value of the \emph{indirect} band gap of 6.36-6.66 eV,\cite{Klimes2014,Jiang2016}
which is only fortuitously in close agreement with experiments.
As a matter of fact, a gap obtained from optical measurements should be compared more appropriately to the smallest \emph{direct} gap in the QP band structure, which in the GWA instead amounts to 11.29--11.66 eV.\cite{Klimes2014}
Moreover, in an optical excitation the electron-hole attraction may lead to the formation of bound excitons, which in turn reduces the optical gap with respect to the QP direct gap. However, the solution of the Bethe-Salpeter equation (BSE), which takes into account excitonic effects accurately,\cite{onida_electronic_2002} has resulted into an exciton binding energy of only 0.35 eV,\cite{satta_many-body_2004} i.e. way too small to resolve the disagreement with the measured optical gap. 

This difference of $\sim 5$ eV between experiments and theory is much larger than it was in the case of \hbn, where the nature and the value of the band gap also remained elusive for a long time. Both experimental and theoretical studies had initially provided underestimated values for the \hbn optical gap, 
mistaking deep-defect-related emissions with the optical band edge.\cite{zunger_optical_1976, taylor_observation_1994}
Cathodoluminescence (CL) measurements in the UV range on high-purity crystals have been achieved only in 2004\cite{watanabe_direct-bandgap_2004} 
and in 2006 GW-BSE calculations were able to reconcile theory and experiment by showing that \hbn is characterized by an indirect QP band gap 
and a large exciton binding energy.\cite{arnaud_huge_2006,Wirtz2006}
Very recently, Cassabois and coworkers, on the basis of photoluminescence (PL) experiments, have corroborated that conclusion, highlighting the role of phonons in the high-energy region of the spectrum and the presence of a weak zero-phonon excitonic line at 5.95 eV.\cite{cassabois_hexagonal_2016,PhysRevB.93.035207}

While the GW-BSE theoretical approach has been proven to be accurate and predictive in \hbn,\cite{arnaud_huge_2006,fugallo_exciton_2015}
there is no reason to expect that it should instead fail for \cbn.
An indication in this direction comes from the good agreement 
between non-resonant inelastic X-ray scattering (NRIXS) spectra on \cbn crystals and GW-BSE calculations of the dynamic structure factor
as a function of momentum transfer $\bfq$.\cite{Galambosi-01}
However, unfortunately NRIXS experiments cannot access the optical limit $\mathbf{q} \rightarrow 0$, making it difficult to determine the optical gap.

The large discrepancy between optical spectroscopy and theory  in \cbn, on the one hand, could be explained by the fact that the gap is most likely too wide to be accessed by standard optical excitation techniques, such as optical absorption and PL. Even using CL, where higher excitations can be in principle addressed, the limitation comes from an efficient detection of the far UV radiation. On the other hand, the generally poor quality of the investigated crystals makes one wonder whether the existing optical spectra are influenced by possible \hbn inclusions \cite{satta_many-body_2004,cappellini_optical_2001,Widmayer1999} and whether they are even largely dominated by defect-related signatures.\cite{tkachev_cathodoluminescence_1985, shipilo_effect_1986, zaitsev_cathodoluminescence_1986,shipilo_influence_1988,  shishonok_near-threshold_2002, manfredotti_vibronic_2006} 

This highly unsatisfactory situation thus calls for a new joint experimental and theoretical effort in order to obtain a definite answer to the issue of the \cbn gap. Furthermore, by analogy with \hbn, interesting optical properties may be expected from \cbn luminescent defects themselves.
Indeed, in the last years it has been demonstrated that several emissions observed within the \hbn optical gap may act as very stable and bright single-photon sources (SPS) in the visible and UV spectral range.\cite{tran_quantum_2016,tran_robust_2016,shotan_photoinduced_2016,jungwirth_temperature_2016,chejanovsky_structural_2016,Martinez_efficient_2016,	exarhos_optical_2017, bourrellier_bright_2016}
Similar sources could be present also in \cbn also in virtue of its  structural and electronic analogies with diamond and silicon carbide, where SPS have been extensively documented.\cite{aharonovich2016solid}

To address the open questions on the optical properties of \cbn it is necessary to investigate absorption and emission spectra in the far UV spectral range. On the one hand, the relatively high density of defects, even in the highest-quality samples, imposes a high spatial resolution to the spectroscopy techniques employed. Conventional optical methods are diffraction limited and even in a confocal microscope a spatial resolution below the hundred of nanometers can hardly be achieved. On the other hand, high-frequency lasers or doubling-frequency techniques are required to excite high-energy levels. 
Both these limitations can be overcome by using high-energy electrons as the exciting radiation. Indeed, in a scanning transmission electron microscope (STEM) electron beams can be focused into sub-nanometric or even sub-\AA ngstrom probes providing the spatial resolution necessary for the study of highly heterogeneous materials. The response function of \cbn can then be accessed in an extended energy range through low-loss electron energy-loss spectroscopy (EELS). Furthermore, the recent development of nano-cathodoluminescence techniques (nano-CL) within a STEM  allows one to investigate the luminescence of a material with a nanometric spatial resolution.\cite{Zagonel2010,KOCIAK2017112} Unlikely photoluminescence, high-energy electrons permit an easy access to an energy range whose upper limit is solely defined by the light collection-analysis optical chain and not by the energy of the exciting radiation.\cite{Ponce1996,bourrellier_nanometric_2014}  Finally, nano-CL provides the capability to address individual color centers and this signal can then be driven to a light-intensity Hanbury-Brown-Twiss  (HBT)  interferometer.\cite{HBT} The potential nonclassical nature of highly 
localized emissions can be demonstrated by the appearance of an anti-bunching signature in the second-order correlation function, $g^{(2)}(\tau)$. Previously, this original experimental setup has been successfully applied in the study of NV centers in nanodiamonds and defect centers in h-BN.\cite{tizei_spatially_2013,PSS2013,bourrellier_nanometric_2014,meuret_photon_2015}

In this work we have investigated the optical properties of \cbn crystals using an experimental and theoretical approach combining electron energy-loss spectroscopy, nano-cathodoluminescence and many-body perturbation theory. In order to probe the intrinsic optical properties of \cbn and to separate defect-related optical features, we have applied these high spatially resolved techniques to the highest-quality samples available.

The paper is organized as follows. In Sec. \ref{sec-Methods} we provide an overview of the theoretical and experimental methods employed. In Sec. \ref{EELSsection} we address the question of the optical  gap of \cbn by EELS. The use of a highly monochromated electron beam permits also the observation of optical phonon modes in the infrared-energy domain of the spectra. These experimental results are then interpreted on the basis of numerical simulations and excitonic and plasmonic features in the loss spectra are identified. In Sec. \ref{annealed-cBN} we show modifications in EELS and nano-CL spectra occurring after high-temperature annealing which promotes local phase changes within the \cbn crystals. In Sec. \ref{sec-defects-emissions} we discuss defect-related emission in \cbn. After an overview of the literature in the field (Sec. \ref{sec-literature-survey}), we present nanometric resolved hyperspectral emission maps of defect emissions (Sec. \ref{sec-nanoCL-exp}). Finally, intensity interferometry experiments have been conducted for the higher-energy features  in order to establish the lifetime of the associated excited state and their possible behavior as single-photon emitters (SPE).

\section{Methods}\label{sec-Methods}
\subsection{Samples preparation}

Colorless high-quality \cbn single crystals have been synthesized by spontaneous nucleation at high pressure and high temperature using the temperature gradient method in \ce{Ba3B2N4} solvent.\cite{taniguchi_spontaneous_2001,taniguchi_synthesis_2007} The crystals have been washed in acetone and ethanol and successively crushed in a mortar to obtain a high-quality submicrometric powder. TEM grids have been prepared by drop casting the crystal dispersed in isopropanol. Lower-quality commercially available micrometric powders (Element Six) have been employed in CL experiments of defect centers. The latter samples have been used also for the study of temperature-induced phase transformations. 

\subsection{Electron energy-loss spectroscopy and cathodoluminescence}\label{sec-CLmethod}

Monochromated EELS experiments were performed on the ChromaTEM microscope  (modified Nion Hermes 200) operated at 60 keV with the sample at room temperature. The electron-beam convergence half angle was 25 mrad and the EEL spectrometer collection half angle was set to about 30 mrad. The electron beam was monochromated to 20 meV full width at half maximum (FWHM)  with an electron-beam current of the order of a few pA. The spectrometer dispersion was set to 1.9 meV/pixel, 9.6 meV/pixel and 53.3 meV/pixel dependently upon the spectral range investigated. Plural scattering contributions were removed using the Fourier-log method.

All nano-CL experiments and the EELS experiments correlated with nano-CL were performed in a dedicated VG-HB501 STEM operating at 100 keV.  The VG microscope was provided with a liquid-nitrogen cooling system for the sample stage (150 K). Electron energy-loss hyperspectral images have been collected using a spectrometer dispersion of 0.1 eV. The spectra have been deconvoluted  using a Richardson-Lucy routine\cite{gloter_improving_2003} after removal of plural scattering. CL signals were collected using an Attolight M\"onch 4107 STEM-CL system fitted with an optical spectrometer with a 300-groove diffraction grating blazed at 2.5 eV or 4.1 eV dependently from the signal analyzed, assuring an energy dispersion of 0.34 nm and 0.16 nm, respectively. The sampling of the spectrometer CCD was 0.17 nm/pixel. Hyperspectral images were obtained by sequentially recording one full CL spectrum per pixel while scanning the sample with a nanometric step size. 

 \subsection{Hanbury Brown and Twiss interferometry}

In our experimental nano-CL setup, the CL signal collected by the parabolic mirror has been driven to a HBT interferometer using an optical multi-mode fiber (\SI{600}{\micro\meter} diameter core). Experiments have been performed in the visible-UV region above 2.59 eV, selected by the convergent lens at the entrance of the optical path (no additional filter has been employed). 
Photon detection has been performed by two photomultiplier (PMT) modules. Time-delay histograms (which are proportional to the second-order correlation function) have been acquired using a Time Harp correlation electronics from Picoquant. The typical room background noise varied from 200 to 500 count/s per PMT. Time-delay histograms have been normalized to one for $\tau\gg0$. 
	
\subsection{Theoretical framework and computational details}

Calculations have been carried out within the framework of many-body perturbation theory.\footnote{For an extended introduction to the theoretical formalism see Ref. \citenum{Martin2016}.}
In the $G_0W_0$ scheme,\cite{Hybertsen1986,Godby1988} the GW self-energy $\Sigma_{xc}$ is obtained as a convolution in frequency space of the Kohn-Sham (KS) Green's function $G_0$ and the dynamically screened Coulomb interaction $W_0$ evaluated in the random-phase approximation (RPA). The QP energies $E_{n\bfk}$ that form the QP band structure are calculated evaluating the first-order perturbative corrections with respect to the KS eigenvalues $\varepsilon_{n\bfk}$:
\beq 
 E_{n\bfk}=\varepsilon_{n\bfk} + Z_{n\bfk}[\text{Re} \Sigma_{n\bfk}^{xc}(\varepsilon_{n\bfk})-V^{xc}_{n\bfk}]
\label{eq:qp2}
\eeq 
with $V_{xc}$ the exchange-correlation KS potential and the renormalization factor $Z$ defined as:
\beq
Z_{n\bfk}^{-1} =  1 - \left. \frac{\partial \text{Re} \Sigma_{n\bfk}^{xc}(\w)}{\partial\w}\right|_{\w=\varepsilon_{n\bfk}}.
\label{eq:zed}
\eeq

In order to calculate the loss function, the BSE two-particle Hamiltonian $H_{\textrm{exc}}$ has to be inverted, which can be formulated as an eigenvalue problem. The direct electron-hole interaction derived from the GWA is $W_0$. Here we use a static approximation to $W_0$. In the  
 one-particle transition basis  $\ket{n}=\ket{\bfk vc} = \phi_{v\bfk}(\bfr)\phi_{c\bfk}(\bfr)$ 
[where $\bfk$ is in the first Brillouin zone and in the Tamm-Dancoff approximation $v$ ($c$) runs over valence (conduction) bands] $H_{\textrm{exc}}$ reads:\cite{Albrecht1998,Benedict1998}
\beq
\bra{n} H_{\textrm{exc}} \ket{n'}  = (E_{c{\bfk}}-E_{v{\bfk}})\delta_{n,n'} + \bra{n} \bar{v}_c-W_0 \ket{n'}.
\label{eqBSE}
\eeq
Here $\bar{v}_c$ is a modified Coulomb interaction in which the long-range component $\bar{v}_c(\bfG=0)$ is set to 0 in reciprocal space. From the eigenvalues $E_{\lambda}$ and eigenvectors $A_{\lambda}$ of the excitonic Hamiltonian \eqref{eqBSE}, one obtains the macroscopic dielectric function $\epsilon_M$, which in the dipole limit $\bfq\rightarrow0$ is:
\begin{multline}
\epsilon_M(\w) =  1 - \lim_{\mathbf{q} \rightarrow 0}  \frac{8\pi}{q^2} \sum_\lambda \Big|\sum_{{\bfk}vc} A_\lambda^{{\bfk}vc}
\bra{\phi_{v\bfk}}e^{-i\mathbf{q}\mathbf{r}}\ket{\phi_{c\bfk+\bfq}}
 \Big|^2 \\
\times \left[\frac{1}{\w-E_\lambda + i\eta}-\frac{1}{\w+E_\lambda + i\eta}\right].
\label{spectrumBSE}
\end{multline}
Here both the resonant and antiresonant contributions (the first and second term in the square bracket, respectively) are explicitly taken into account, which is important for EELS spectra calculations.\cite{Olevano2001} The effect of the electron-hole interactions $\bar{v}_c$ and $W_0$  in  Eq. \eqref{eqBSE} is the mixing of the amplitudes through the coefficients $A_\lambda$ in the first line of \eqref{spectrumBSE} and the change of the excitation energies $E_\lambda$ in the denominators of the second line, which in absence of excitonic effects would just correspond to energy differences between QP bands: $E_{c{\bfk}}-E_{v{\bfk}}$.

Finally, while $\text{Im} \epsilon_M(\w)$ yields the absorption spectrum,
the loss function $-\text{Im} \epsilon_M^{-1}(\w)$ measured by EELS (for $\bfq\rightarrow0$) can be conveniently expressed as:
\beq
-\text{Im} \epsilon_M^{-1}(\w) = \frac{ \text{Im} \epsilon_M(\w)} { [\text{Re} \epsilon_M(\w)]^2 + [\text{Im} \epsilon_M(\w)]^2}.
\label{eqloss}
\eeq
Peaks in the loss function derive either from $\text{Re} \epsilon_M =0$ with $\text{Im} \epsilon_M$ not too large, which corresponds to plasmon resonances,
or from peaks of $\text{Im} \epsilon_M$, which corresponds to (screened) interband transitions. 

In this work we have employed norm-conserving Troullier-Martins pseudopotentials\cite{PhysRevB.43.1993} in a plane-wave approach.\cite{abinit}
KS eigenenergies and eigenfunction have been obtained in the local-density approximation (LDA) with a cutoff of 30 Hartree, adopting the experimental lattice parameter $a=3.615$ \AA.\cite{Wyckoff} The Brillouin zone has been sampled with a 4-times shifted 8$\times$8$\times$8 Monkhorst-Pack grid of  $\bfk$ points (which for the calculation of spectra were shifted to avoid high-symmetry points). In the $G_0W_0$ calculations we have used the Godby-Needs plasmon-pole approximation,\cite{PhysRevLett.62.1169}
including 300 bands in the evaluation of the self-energy.
The resulting GW corrections to the LDA were taken into account in the BSE by applying a scissor operator of 1.86 eV to open the LDA  band gap and by stretching valence and conduction bands by  6.7 \% and 4.5 \%, respectively.
The BSE spectra converged with 30 bands. In the reciprocal space the dielectric function was represented as a matrix of size 89 $\mathbf{G}$ vectors.
To compare the calculated loss function with the experiments we introduced a 0.5 eV gaussian broadening. For BSE calculations we used the EXC code.\cite{exc}

\section{Results and discussion}

\subsection{Electron energy-loss spectroscopy}\label{EELSsection}

Low-loss EELS is a very powerful technique to investigate material spectral response in the wide energy-range relevant for large band-gap materials. The few EELS spectra in the UV range published on poor quality \cbn crystals presented a very poor energy resolution and a sole broad electronic excitation centered at about 30 eV could be identified.\cite{schmid_phase_1995,Weinmantel-2002} 

\begin{figure}[t!]

\includegraphics[width=\columnwidth]{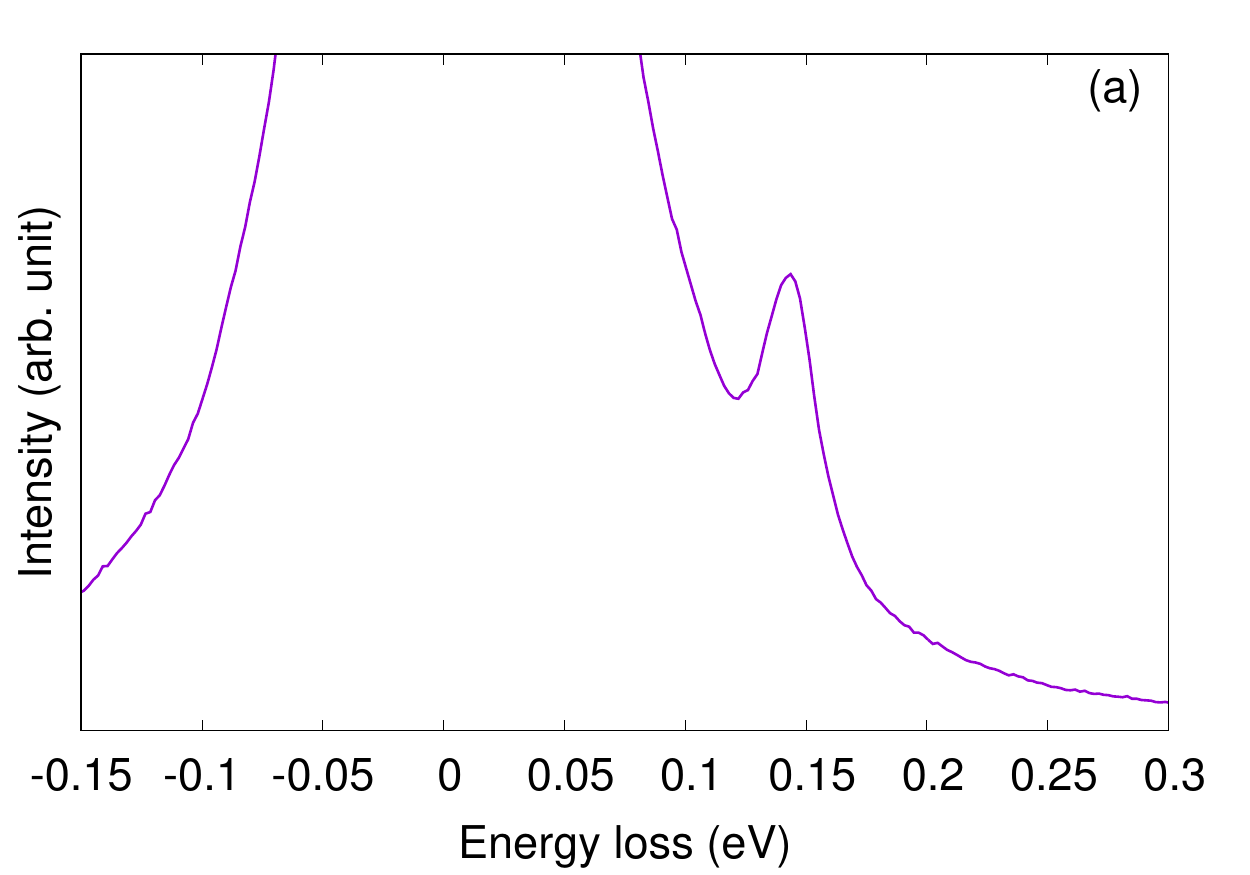}

\includegraphics[width=\columnwidth]{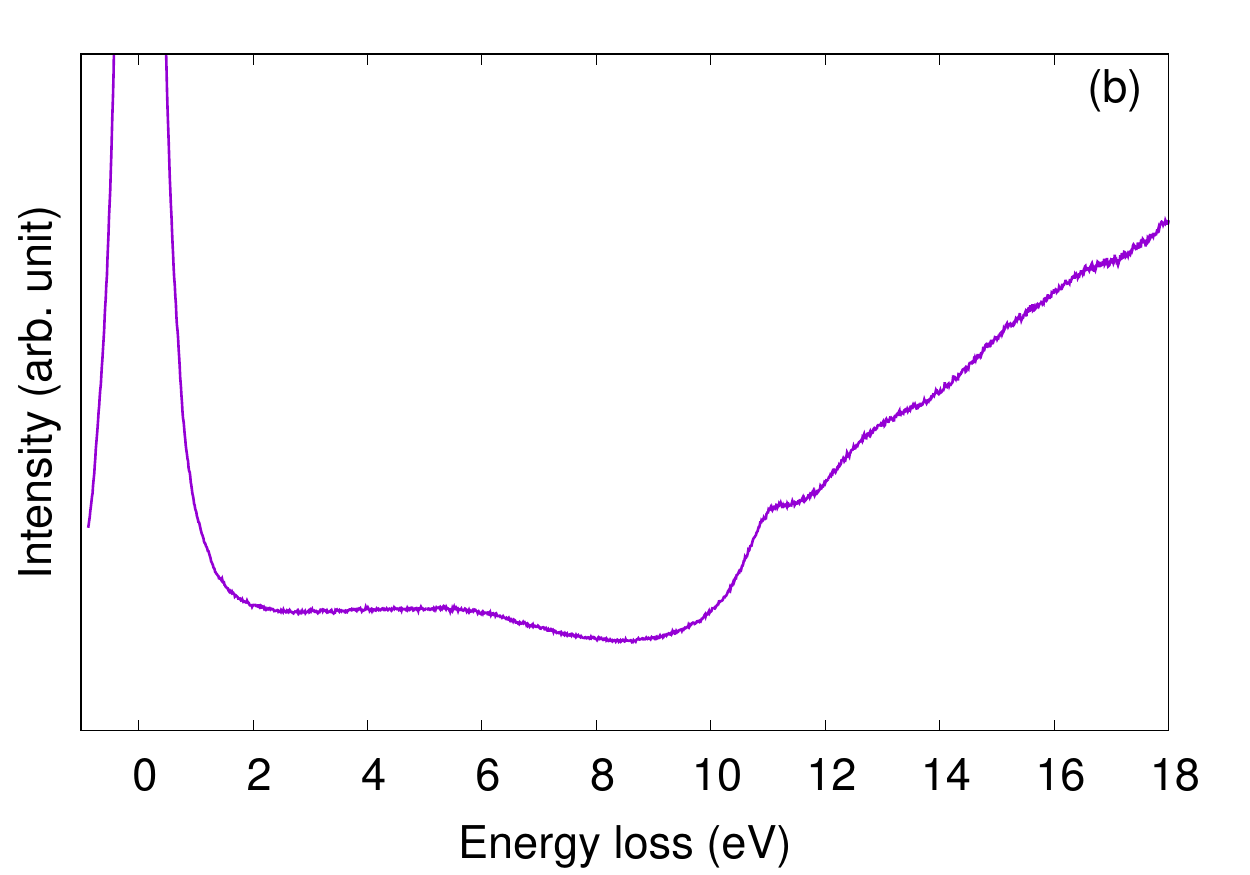}

\includegraphics[width=\columnwidth]{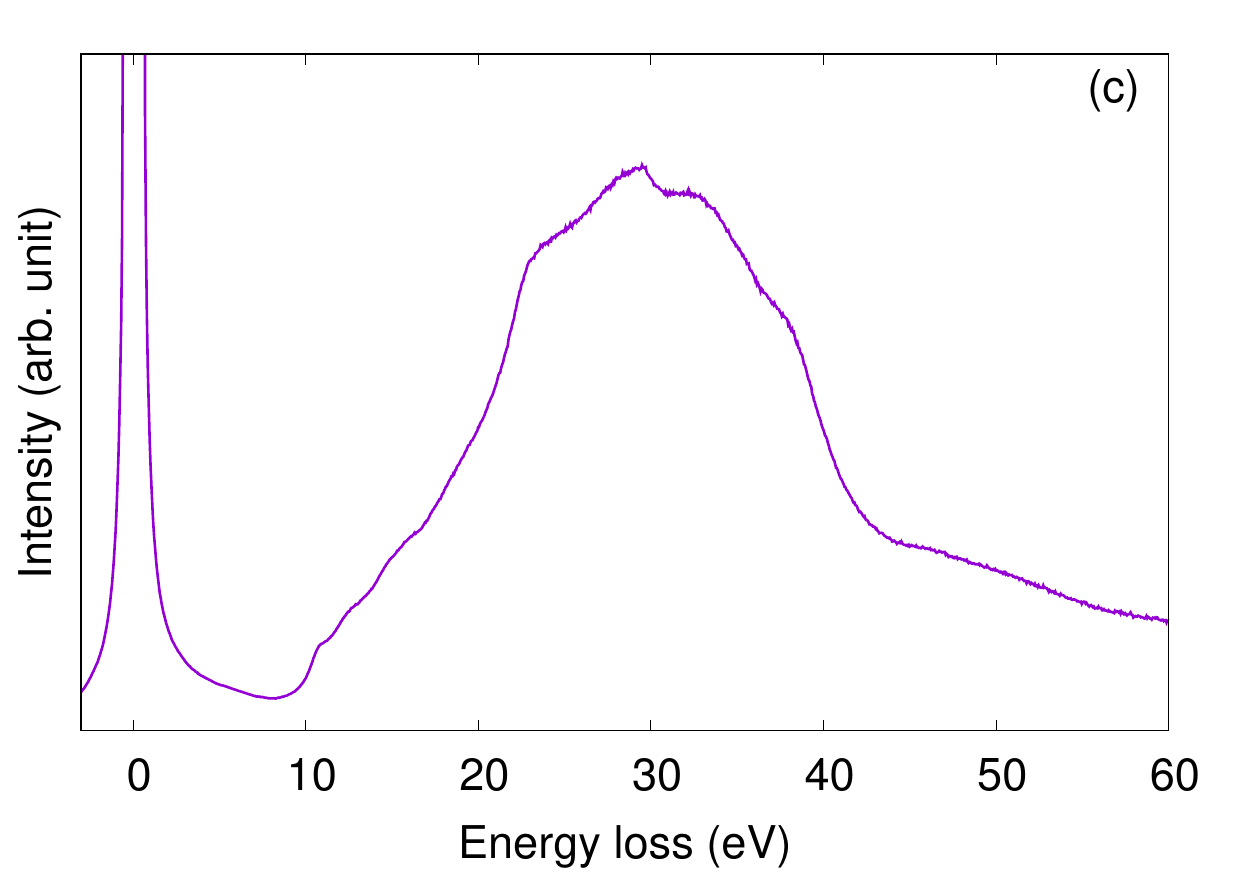}
\caption{EELS low-loss full spectrum of a thin region (about 80 nm thick) of a \cbn high purity crystal. (a) Infrared spectral region showing a prominent peak corresponding to \cbn optical phonon modes. (b) Spectral region focusing on the onset of the response function. The weak  signal in the lower energy region corresponds to the signature of Cherenkov effects. (c) Complete loss function presented after removal of multiple scattering.} \label{full-spectrum}
\end{figure}

In Fig. \ref{full-spectrum} we present three low-loss EELS spectra acquired from the same area of a 80 nm thick \cbn crystallite derived by crashing a high-quality macroscopic crystal. The thickness $t$ has been estimated from the Poisson statistics as $t/{\lambda}=\ln(I_t/I_0)$, where $I_t$ is the total area of the spectrum, $I_0$ is the area under the zero-loss peak (ZLP) and $\lambda$ is the mean free path for inelastic scattering, derived from acquisition parameters and the effective atomic number of the specimen.\cite{egerton_electron_2011} Thinner regions of the crystals were not considered to minimize surface-loss contributions to the spectra. In Fig. \ref{full-spectrum}(a) a well defined peak appears at 143 meV which corresponds to optical phonon modes of \cbn, in accordance with previous EELS experiments.\cite{cBN-phonons-Ramasse-2015,nicholls2018theory,Hageeaar7495} 
As the experiments are performed on the bulk of a cBN crystal and are not resolved in momentum, we expect in an EELS experiment to preferentially measure the longitudinal-optical  mode (LO) with respect to the transverse-optical (TO). In Fig. \ref{full-spectrum}(b) the onset can be clearly identified with the rise of the loss function around 10.1 eV and the appearance of a prominent peak at 11.1 eV. The small constant background measured within the forbidden optical region which progressively decreases at 6 eV can be tentatively attributed to Cherenkov or Transition radiations (CR or TR) effects. CR happen when the index of the medium is larger than the speed of light to electron speed ratio. In our case, the calculated static index ($\sim 2.16$) is slightly smaller than the light to electron speed ratio ($\sim 2.24$) which suggests that the small plateau most probably arises from TR.
The low-loss spectrum recorded in a wider energy range is presented in Fig. \ref{full-spectrum}(c) where a broad and structured peak centered at 33 eV is visible. These results are consistent with the NRIXS spectra\cite{Galambosi-01} even though the present EELS experiment is carried out at the optical limit, i.e. with a momentum transfer $\mathbf{q} \rightarrow 0$, while  NRIXS experiments were performed at much higher $\mathbf{q}$. 

\begin{figure}[bt]
	\centering
	\includegraphics[width=\columnwidth]{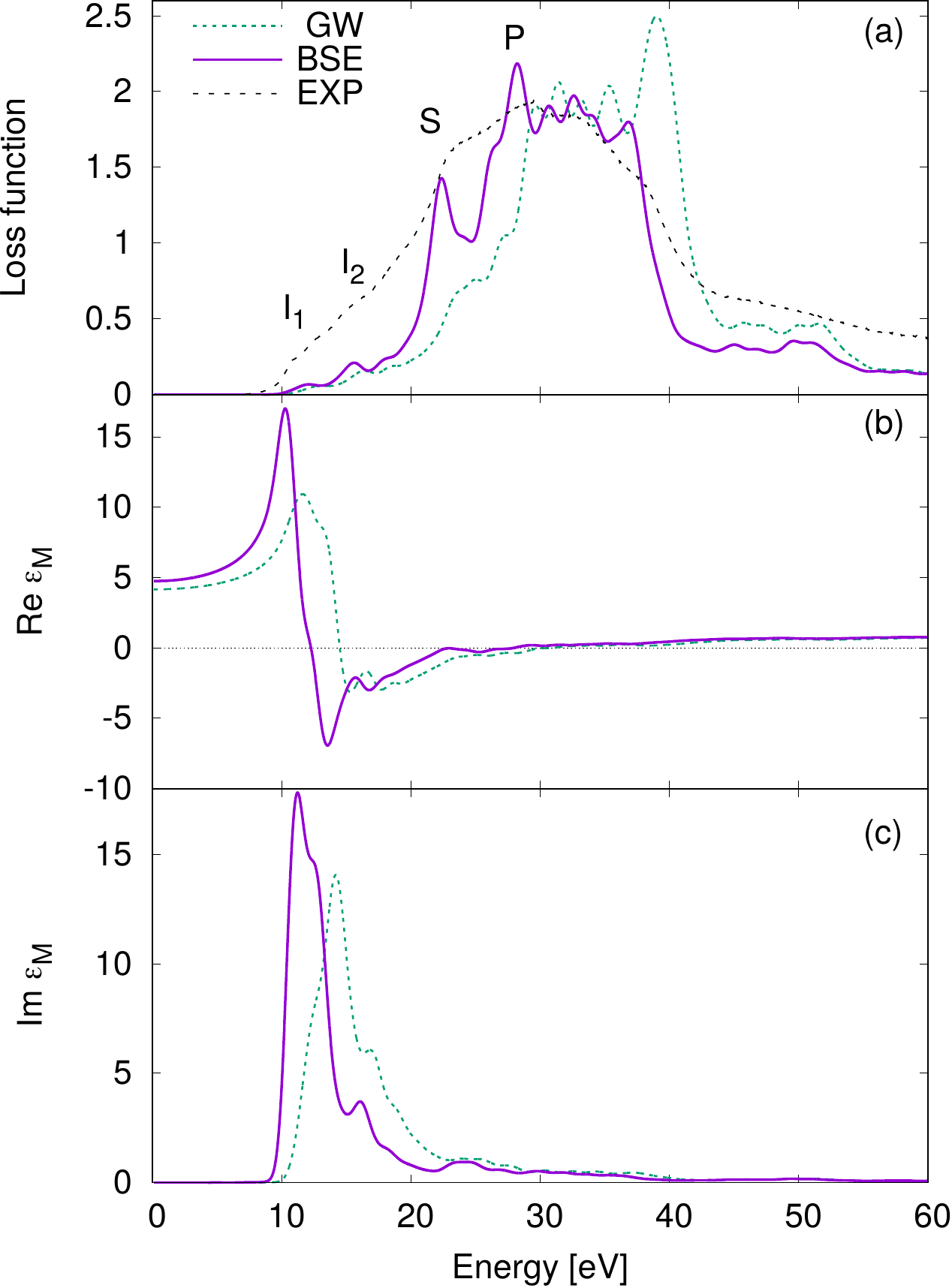}
	\caption{(a) The loss function, (b) the real and  (c) the imaginary part of the dielectric function calculated from the BSE (violet solid lines) and neglecting the electron-hole interactions in the GW approximation (green dashed lines). In (a) the calculated loss functions are compared with the measured spectrum where the zero loss peak has been removed (black dashed line).}
	\label{spectra}
\end{figure}

\begin{figure}[bt]
	\centering
	\includegraphics[width=\columnwidth]{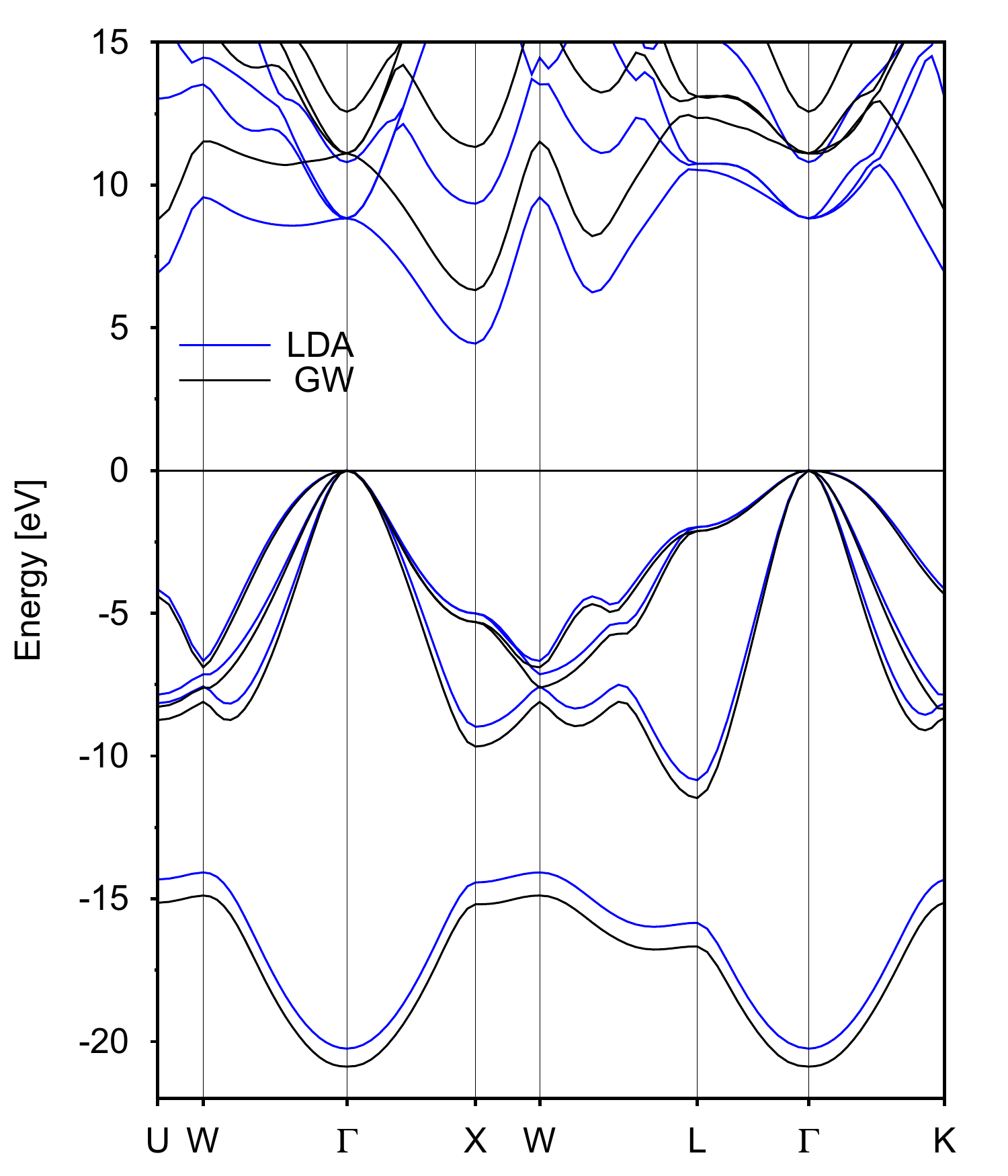}
	\caption{Band structure of c--BN from KS eigenvalues within the LDA (blue lines) and from the GWA in the $G_0W_0$ approach (black lines). The zero of the energy axis has been set to the top-valence maximum.
		\label{bands}} 
\end{figure}

In order to interpret these experimental findings, in Fig. \ref{spectra}(a) we compare the measured EELS spectrum where the zero-loss peak has been removed (black lines) with the loss function calculated from the BSE (violet solid lines).
The overall good agreement allows us to analyse the origin of the different spectral features  on the basis of the calculations. The main broad peak P is the valence plasmon resonance:  it corresponds to a zero of $\text{Re} \epsilon_M$, see Fig.  \ref{spectra}(b). The onset of the calculated loss function is characterized by the two smaller structures I$_1$ and I$_2$.
They are interband transitions that stem from the  peaks in the absorption spectrum $\text{Im} \epsilon_M$,  see Fig.  \ref{spectra}(c), and are only slightly blueshifted by the contribution of  $\text{Re} \epsilon_M$ to  Eq. \eqref{eqloss}. The onset of the calculated spectra has an excitonic origin. 
This can be easily understood by comparing the BSE spectra (violet solid lines) with the GW spectra (green dashed lines),  where the electron-hole interactions in Eq. \eqref{eqBSE} have been neglected: besides a small redshift of all the spectra, as a result of excitonic effects both the real and imaginary parts of the dielectric function get strongly enhanced towards lower energies. In the GW absorption spectrum the smallest-energy transitions take place between the threefold-degenerate top-valence and bottom-conduction bands at the $\Gamma$ point of the band structure, see Fig. \ref{bands}. The minimum direct gap at $\Gamma$ is  8.83 eV in LDA (blue lines in Fig. \ref{bands}) and becomes 11.10 eV in GW (black lines in Fig. \ref{bands}). The effect of the GW corrections in the band structure is mainly a constant opening of the band gaps along the different $\bfk$ points, justifying the use of a scissor operator and a small stretching of the bands. The electron-hole interaction in the BSE gives rise to nine degenerate bound excitons in the absorption spectrum, with a  binding energy of $\sim 0.3$ eV. We note that the binding energy of the excitons in \cbn is smaller than in \hbn, where it is 0.7 eV.\cite{arnaud_huge_2006} Indeed, while in \hbn the bound excitons derive from transitions between parallel flat bands, supporting an interpretation in terms of a tightly bound Frenkel exciton,\cite{arnaud_huge_2006} in the case of \cbn the bands at the origin of the bound excitons are parabolic, leading to a Wannier-exciton picture. \cite{satta_many-body_2004} The calculated BSE optical gap in the absorption spectrum thus amounts to $\sim 10.7$ eV,  i.e. much larger than the  $\Gamma$X  indirect band gap, which is 6.31 eV in GW (4.45 in LDA). Excitonic effects are important also for the rest of the spectrum, shifting the spectral weight towards lower energies and improving considerably the agreement with the experimental spectrum. Notably, the shoulder S on the low-energy side of the main plasmon peak P in the loss function is quite weak in the GW spectrum and gets strongly enhanced, in agreement with experiment, when electron-hole interactions are taken into account in the BSE.

The values of the $G_0W_0$ quasiparticle gaps in the band structure and the optical gap in the BSE absorption spectrum agree well with the corresponding results from recent calculations.\cite{Klimes2014,Jiang2016,satta_many-body_2004}
By using a more accurate partially self-consistent GW approach  instead of the perturbative $G_0W_0$ scheme employed here, it has been shown that the direct  gap in the quasiparticle band structure would be even larger by 0.3 eV.\cite{Klimes2014}

\subsection{Detecting phase transformations in annealed \cbn}\label{annealed-cBN}

Boron nitride exists in two low-density $sp^2$ phases, the hexagonal and rhombohedral (\rbn) allotropes, where weakly interacting planes of hexagonally arranged B and N atoms are stacked respectively in a AA$^{\prime}$ and ABC order. The $sp^3$ coordination gives rise to two further high-density phases,  the cubic and wurtzite (\wbn) allotropes. The latter structure corresponds to a hexagonal lattice isoelectronic with the carbon allotrope lonsdaleite. The stacking similarities between the different BN structures suggest that the conversion of \hbn into \wbn and \rbn into \cbn can be driven by simple compression, through the formation of new out-of-plane covalent bonds.\cite{mirkarimi_review_1997, nistor_crystallographic_2005} On the contrary, the \hbn to \cbn transition requires bond breaking and atom diffusion to transform the AA$^{\prime}$ stacked \hbn atomic planes into the ABC ordered puckered lattice planes in the $\left\langle 111\right\rangle$ direction of c-BN face-centered cubic lattice. HRTEM studies have reported evidences against a direct \hbn to \cbn transformation, in favor of the formation of transient phases mediating the conversion.\cite{horiuchi_monoclinic_1996, nistor_crystallographic_2005} Single and multiwalled nanoarches formed by joining dangling bonds at \hbn sheet edges have been observed in the starting \hbn material.\cite{nistor_crystallographic_2005} At the beginning of the high-pressure/high-temperature conversion, compression and local folding of \hbn planes occur, inducing a transition to a monoclinic crystal structure.\cite{horiuchi_monoclinic_1996,	nistor_crystallographic_2005} Moreover, additional nanoarches form and are believed to act as nucleation sites for c-BN crystallites [the nanoarches planes becoming the (111) planes of the cubic lattice].\cite{nistor_crystallographic_2005} The coalescence of pairs of planes promoting the stacking order change is associated with an \rbn arrangement. In small \cbn grains, \wbn areas have been observed \cite{horiuchi_monoclinic_1996}.

These findings demonstrate the relative abundance in \cbn of defects consisting in inclusions of hexagonal-related phases. Besides these native defects, it is possible to deliberately induce the coexistence of several phases in higher-quality \cbn crystals by promoting inverse transformation. The phase-change temperature from \cbn to \hbn at ambient pressure has been experimentally measured at $1320^{\circ}\text{C}\pm 380^{\circ}\text{C}$,\cite{WILL2000280} in good agreement with the \hbn$\rightleftharpoons$\cbn equilibrium line proposed in the theoretical phase diagram of BN.\cite{Solozhenko-99} By annealing \cbn at sufficiently high temperature, the first stages of this structural transformation correspond to the nucleation of tiny hexagonal domains. This local phase change could in principle be optically detected using spatially resolved techniques such as EELS and CL. 

\begin{figure}[tb]
	\includegraphics[width=\columnwidth]{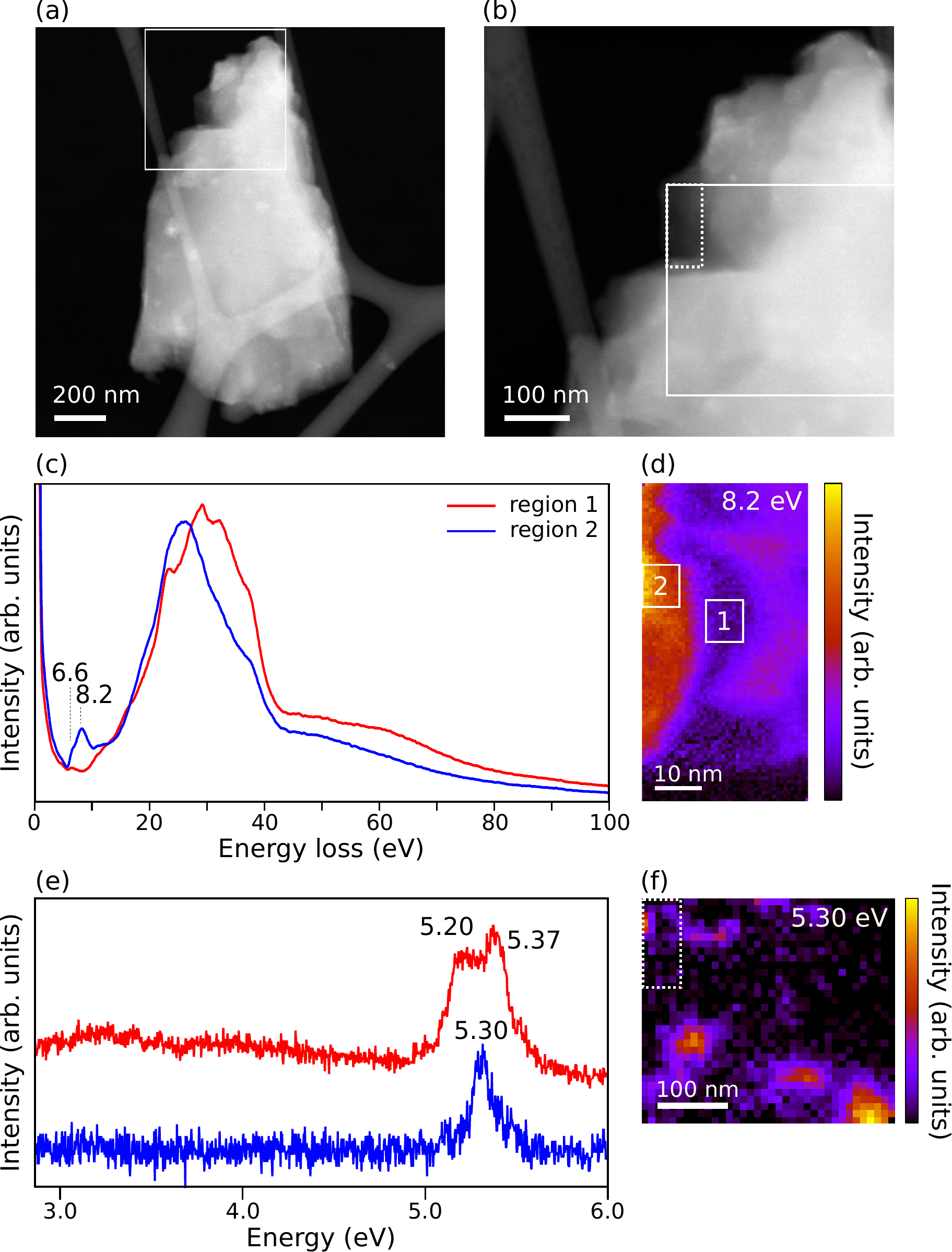}
	\caption{(a),(b) HAADF images of a micrometric particle from the thermally treated $c$--BN powder. The smaller rectangle in (b) indicates the region investigated by EELS hyperspectral imaging, the larger rectangle the one investigated by CL hyperspectral imaging. (c) EELS spectra extracted from regions 1 and 2 indicated in the intensity map in (d). The color scale in (d) represents the intensity of the 8.2 eV feature. (e) Typical CL signals occurring at the thermal treated $c$--BN powder. CL maps obtained integrating the luminescent signal in the 4.8--6.0 eV energy window. \label{c-BN-1200C}} 
\end{figure}

To prove this, commercially available micrometric powders have been heated in a tubular furnace at 1200~$^{\circ}$C for 36 hours in an inert atmosphere. In Fig. \ref{c-BN-1200C}(a) we present high-angle annular dark-field (HAADF) images of a particle obtained after crashing the treated powder in a mortar and in Fig. \ref{c-BN-1200C}(b) a magnified image of the region indicated by the rectangle in Fig. \ref{c-BN-1200C}(a). The morphology of the particle does not indicate any significant difference compared to the untreated system. An EELS hyperspectral image has been acquired in the area marked by the small dashed rectangle in Fig. \ref{c-BN-1200C}(b). The region has a rather constant thickness as deduced by the homogeneity of the HAADF image intensity. However, the intensity of the signal integrated after the ZLP subtraction in a 0.7 eV window centered at 8.2 eV  presents strong spatial variations [Fig. \ref{c-BN-1200C}(d)]. Indeed, EELS spectra are fundamentally identical to those of untreated \cbn powders almost everywhere in the particle [red spectrum in Fig. \ref{c-BN-1200C}(c)] but for two pronounced features (6.6 eV and 8.2 eV) which appear within the \cbn band gap [blue spectrum in Fig. \ref{c-BN-1200C}(c)] at specific positions in the particle. Furthermore, the higher-energy region appears modified, with a dominant contribution at about 26.0 eV. These additional components match the energy of the main plasmon peaks of the h-BN 
bulk spectrum.\cite{arenal_electron_2005, tarrio_interband_1989}

A further confirmation of a local phase transformation can be derived by an 
analysis of the cathodoluminescent signal. In Fig. \ref{c-BN-1200C}(e) we 
present spectra extracted from a CL hyperspectral image acquired in a wider area of the crystal overlapping the previous EELS analysis zone [corresponding to the solid rectangle in Fig. \ref{c-BN-1200C}(b)]. A larger analysis region is now possible since cathodoluminescence does not require the thin-sample constrain as for EELS. Compared to the luminescence of the pristine sample, we observe the appearance of peaks at 5.20 eV, 5.30 eV and 5.37 eV. The map of the 5.30 eV emission peak [Fig. \ref{c-BN-1200C}(f)] presents very localized emission spots with a FWHM size of just few tens of nanometers, i.e. of the same order of magnitude of defect emissions later discussed in Sec. \ref{sec-nanoCL-exp}. Notably, this emission spot occurs in the area previously analyzed by EELS (dashed rectangle) and it is spatially correlated with the additional features observed by EELS.

These spectroscopic observations are strong indications for a local phase 
transformations within the particles. EELS spectra suggest the presence of \hbn
domains but the cathodoluminescence spectra do not show the characteristic 
5.75 eV emission of the pure \hbn phase. Previous studies have demonstrated 
that the CL spectrum of \hbn, unlike EELS spectra, is strongly affected by 
changes in the stacking order.\cite{bourrellier_nanometric_2014} One may guess 
that these domains present a hexagonal phase which does not respect the AA$^{\prime}$ stacking order of bulk \hbn. As we discussed at the beginning of this section, \cbn can be obtained by a direct transformation from the \rbn phase and only indirectly from the \hbn phase. Therefore, the inverse transformation would drive to the \rbn phase having an ABC stacking. From another side, it has been demonstrated that \cbn presents stacking faults that can be described as very local \wbn phase domains.\cite{horiuchi_monoclinic_1996} In that case, temperature treatments could induce the relaxation of these native local domains to the more stable \hbn. Finally, the non-correspondence on CL spectra could be due to confinement effects. These aspects of the observed phase transformation are open to further investigations and certainly reference spectra of pure \rbn and \wbn would provide an important contribution. However, this spectroscopic information is currently unavailable due to the lack of good-quality crystals for these metastable phases.

\subsection{Optically active defects}\label{sec-defects-emissions}

The very large band gap of \cbn makes it unlikely to be employed as an optical functional material, but its technological potential might reside in the presence of bright and stable optical defects. During the last 30 years, many luminescence features have been observed by PL and CL in a spectral range going from the infrared to the far UV. Studies have been conducted on a variety of \cbn samples such as crystal grown in B--rich or N--rich conditions, annealed, electron irradiated, ion bombarded or compressed. Many of these emissions present a rich vibrational structure with several phonon replica which can be related  with main \cbn phonon modes.\cite{Shishonok-vibrations-2003,Shishonok-vibrations-2007} Whereas complete direct measurements of \cbn phonon band structure have 
not been reported yet, theoretical values obtained in the framework of DFT can be confidently employed to identify well-defined phonon 
modes.\cite{kern_ab_1999, karch_ab-initio_1997, bechstedt_dynamics_2000, 
bagci_ab_2006} Despite a rich literature in the field, no complete characterization (concerning lifetime, spatial localization, polarization, etc.) of these emission lines has been reported yet. Furthermore, \cbn might contain single-photon sources as in diamond or SiC, but this hypothesis, which received also a recent theoretical support,\cite{abtew_theory_2014} has not been explored experimentally. This situation motivates further experimental studies using modern characterization techniques. In this section, after an overview on previous works, we present new results on defect luminescence in \cbn obtained by nano-CL showing the high spatial localization of its defect-related emissions and, for some of them, their lifetime.

\subsubsection{A literature survey}\label{sec-literature-survey}

A rather fragmented picture emerges from the rich literature on defect luminescence in \cbn with a broad and sometimes confusing nomenclature. 
Here we survey earlier studies on the luminescence of \cbn with the aim to provide a unified vision of the field. Key information has been reported in Table \ref{emissions-summary}. Tentative attributions of different emission lines to intrinsic or extrinsic defects have been discussed most of times on the sole correlation between growth conditions or post-growth treatments and luminescence observations. Whereas these hypotheses are often very plausible, the lack of more complete experimental characterizations does not permit an unequivocal identification of the observed defects  and therefore generic references to vacancies, interstitial or impurities have been often employed.

\begin{table*}
\caption{Summary of the major emission lines observed on $c$--BN with 
indication of their lifetime, main phonon replica and tentative attribution indicated using the Kr\"oger-Vink notation where possible.}
 \begin{tabularx}{\linewidth}{lldddLL}
 \hline
 \hline
 Series & Nomenclature & \multicolumn{1}{c}{Energies} & \multicolumn{1}{c}{Lifetime} & \multicolumn{1}{c}{Main phonon}  & Tentative & References\\
  &  &  \multicolumn{1}{c}{(eV)} & \multicolumn{1}{c}{(ns)}&    \multicolumn{1}{c}{replica (meV)} &  attribution & \\
 \hline
 \hline
General cubic & GC-1 & 1.76 & & 64 & N 
vacancies complexes & \citenum{tkachev_cathodoluminescence_1985, 
shipilo_effect_1986}\\ 
& GC-2 & 1.63 & & 56 & B vacancies complexes, $O_N$--$V_B$ & \citenum{tkachev_cathodoluminescence_1985, 
shipilo_effect_1986},this work\\ 
& GC-3 & 1.55 & &  &  & \citenum{tkachev_cathodoluminescence_1985, 
shipilo_effect_1986}\\ 
& GC-4 & 1.44 & & 56 &  & \citenum{shipilo_effect_1986}\\
\hline
 & M1\footnotemark[1] & 1.65 & &   & $C_B$ & \citenum{shipilo1991luminescent,Erasmus-04} \\
 & M2 & 1.64 & &  & $Si_B$ & \citenum{shipilo1991luminescent} \\
 & $\Gamma$ & 2.99 & & 33 &  Stoichiometry changes & \citenum{shipilo1991luminescent}\\
 \hline
Irradiation induced  & RC-1 & 2.27 &  & 64 &  Vacancies& \citenum{zaitsev_cathodoluminescence_1986,Shishonok-04,Erasmus-04,buividas2015photoluminescence},this work\\
defects & RC-2 & 2.15 & & 64 & Vacancies& \citenum{zaitsev_cathodoluminescence_1986,Shishonok-04,Erasmus-04,shishonok_near-threshold_2002,buividas2015photoluminescence},this work\\
& RC-3 & 1.99 & 3.7 & 64 &  Vacancies& \citenum{zaitsev_cathodoluminescence_1986,Shishonok-04,Erasmus-04,shishonok_near-threshold_2002,buividas2015photoluminescence}\\ & RC-4 & 1.86 & & 64 &Vacancies  & \citenum{Shishonok-04,Erasmus-04,shishonok_near-threshold_2002}\\ & BN-1 & 3.3 & & 161 & $B_i$ & \citenum{shishonok_near-threshold_2002,Shishonok-vibrations-2007}\\
\hline
Pressure cubic & PC--1 & 2.84 & & 100 & Interstitial atoms & \citenum{shipilo_influence_1988,Shishonok1989}\\ 
& PC--2 & 2.33 & & 100 & Interstitial atoms & \citenum{shipilo_influence_1988,Shishonok1989}\\ 
& PC--3\footnotemark[2] & 1.79 & &  & Interstitial atoms, transition metals impurities & \citenum{shipilo_influence_1988,Shishonok1989,Shishonok-92, shishonok_near-threshold_2002}\\ 

\hline & PF--1\footnotemark[3] & 3.57 & 0.55 & 140 &  $V_N^{-}$--$B_i^{+}$, $B_N$& \citenum{shipilo_electron-vibrational_1990,
shishonok_near-threshold_2002,mishima_electric_2000, 
zhang_cathodoluminescence_2002,Shishonok-vibrations-2003, 
manfredotti_vibronic_2006},this work \\
\hline
& GB--1\footnotemark[4] & 3.21 & & 25 & Impurities & \citenum{shishonok_luminescence_2007, shipilo1991luminescent}\\

\hline & T & 4.93 & 1.4 & 145 & Si impurities & \citenum{shipilo1991luminescent,kanda1997growth,evans_determination_2008},this work \\
\hline &  & 4.09 & & 175 & $C_N$ at $h$--BN inclusions & \citenum{shishonok_luminescence_2007}\\
\hline
\hline
 \end{tabularx}
  \footnotetext[1]{This emission has been named R center in Ref. \citenum{Erasmus-04}.}
  \footnotetext[2]{This center might be the same as the center named G in Ref. \citenum{Shishonok-92}.}
   \footnotetext[3]{In several works the first phonon replica has been named also PF--2. The full  	band, observed without resolving the phonon structure, has 	also been named UCL, US--1.}
  \footnotetext[4]{This emission has been named O center in Ref. \citenum{shipilo1991luminescent}.}
\label{emissions-summary}
\end{table*}

\begin{itemize}
\item  
The first series of color centers at energies GC--1=1.76 eV, GC--2=1.63 eV, 
GC--3=1.55 eV and GC--4=1.44 eV (GC for General Cubic), was identified in 
1985 by Tkachev et al. by CL on polycrystalline $c$--BN 
materials.\cite{tkachev_cathodoluminescence_1985} These centers present phonon replica with well identified energies, 64 meV for the CG-1 center and 56 meV for the GC--2 and GC--4 centers. Later studies conducted under high-temperature annealing conditions and nitrogen implantation permitted to ascribe the GC--1 center to defects containing nitrogen vacancies and the GC--2 center to defects containing boron vacancies.\cite{tkachev_cathodoluminescence_1985, shipilo_effect_1986} A more recent theoretical work proposed the GC--2 center to be related to substitutional oxygen at a nitrogen site adjacent to a boron vacancy\cite{abtew_theory_2014} ($O_N$--$V_B$), which would be an analogous in \cbn to the well known N--V$^{-}$ center in diamond.

\item 
In the presence of carbon or silicon doping two centers appear at the energies M--1=1.65 eV and M--2=1.636 eV.\cite{shipilo1991luminescent}  The intensity of these centers increases when the growth occurs in a nitrogen-excess atmosphere which might promote the formation of boron vacancies. Therefore, the M--1 center has been ascribed to C or Si substitutionals at B sites ($C_B$) while the M--2 center to Si substitutionals at B sites ($Si_B$). The centers can be seen as an evolution of the GC--2 line in the presence of dopant atoms.
When the growth occurs in a  boron excess, carbon or silicon doping leads to the appearence of the $\Gamma$ center at 2.99 eV presenting a phonon side band at 33 meV.\cite{shipilo1991luminescent}

\item  A series of emissions named Radiation Cubic (RC) has been observed by CL at the energies RC--1=2.27 eV, RC--2=2.15 eV, RC--3=1.99 eV and RC--4=1.86 eV on  $c$--BN crystals bombarded with electrons having an energy in the order of the MeV or in the range 150--300 keV\cite{zaitsev_cathodoluminescence_1986,shishonok_near-threshold_2002,Shishonok-04, Erasmus-04} or by ion irradiation.\cite{Zaitsev-87} More recent investigations reported the generation of these centers by femtosecond laser pulses.\cite{buividas2015photoluminescence} In the same work, a PL decay close to a single exponential with time constant 3.7 ns has been also reported for the RC--3 line. All emissions present phonon replica spaced by 64 meV. These features have been ascribed to vacancy-related defects since they disappear after annealing at temperatures around 1000 $^{\circ}$C.\cite{zaitsev_cathodoluminescence_1986, shishonok_near-threshold_2002} Zero-phonon lines of the RC--1 and RC--2  centers present finer structures, depending upon the particle irradiation energy, which have been associated to local symmetry breakings at the defect centers.\cite{Shishonok-04,Erasmus-04}
Another electron-irradiation induced emission, named BN--1 center, was identified at 3.29 eV accompanied by a very rich phonon structure.\cite{shishonok_near-threshold_2002,Shishonok-vibrations-2007} By analogy with a similar center in diamond, the BN--1 center was thought to include most probably a boron interstitial ($B_i$).\cite{shishonok_near-threshold_2002}

\item  
Emissions observed by CL after high-pressure treatments at almost 10 GPa have been classified as Pressure Cubic (PC) PC--1=2.84 eV, PC--2=2.33 eV and PC--3=1.79 eV. These centers present local phonon replica of 100 meV,\cite{shipilo_influence_1988, shishonok_near-threshold_2002} a higher energy than in the GC or the RC series. This suggested the possible presence of interstitial atoms. An influence of transition metal impurities has been suggested for the PC--3 line.\cite{Shishonok1989}	

\item  A broad emission band centered at about 3.12 eV has been reported by several works and different notations such as UCL or US--1 have been used for 
it.\cite{mishima_electric_2000, zhang_cathodoluminescence_2002, 
manfredotti_vibronic_2006} Other works reported the presence of two sharp peaks 
at the high-energy side of the band, indicated as PF--1=3.57 eV and PF-2=3.41 
eV.\cite{shipilo_electron-vibrational_1990,shishonok_near-threshold_2002} Low-temperature photoluminescence and cathodoluminescence measurements demonstrated a very rich phonon structure for the PF--1 center, the PF--2 center being then the first intense phonon replica. \cite{shipilo_electron-vibrational_1990,Shishonok-vibrations-2003} The broad bands previously discussed in the literature correspond to the unresolved phonon replica. This center has been interpreted as related to donor-acceptor Frenkel pairs, $V_N^{-}$--$B_i^{+}$, by comparing CL of B-implanted, electron-bombarded and annealed 
crystals.\cite{shipilo_electron-vibrational_1990, shishonok_near-threshold_2002} Manfredotti \textit{et al.} proposed a different assignment with a boron substitutional at a nitrogen site ($B_N$).\cite{manfredotti_vibronic_2006}

\item  
A series of five sharp lines spaced by 20-25 meV has been observed in the 
energy range 3.23 and 3.13 eV, with intense peaks at 3.18 eV and 3.23 eV. This series has been called GB--1 center but also O center.\cite{shishonok_luminescence_2007, shipilo1991luminescent} 
Shipilo \textit{et al.} reported that these emissions are encountered in Si or carbon doped crystals only grown in an excess of nitrogen.\cite{shipilo1991luminescent} Shishonok et al. proposed that this center might be associated with heavy atom impurities due to the low energy of the phonon replica which does not correspond to any typical vibration mode of \cbn.\cite{shishonok_luminescence_2007}

\item A zero-phonon line at 4.93 eV followed by several phonon replica 
separated by 152 meV, originally indicated as T center, has been observed using cathodoluminescence\cite{shipilo1991luminescent,kanda1997growth} and synchrotron radiation stimulated luminescence.\cite{evans_determination_2008} This emission has been firstly detected in deliberately highly Si doped crystals and interpreted as the interaction between Si impurities with other defective sites.\cite{shipilo1991luminescent}

\item In lower-quality $c$--BN crystals, a 4.09 eV emission followed by phonon 
replica spaced by 175 meV has been observed.\cite{shishonok_luminescence_2007} 
These features are typical of \hbn, which is a common inclusion in
\cbn crystals, and they have been discussed as being associated to carbon 
substitutionary atoms at nitrogen sites.\cite{katzir_point_1975,taniguchi_synthesis_2007,bourrellier_bright_2016}

\end{itemize}

We have limited this summary to intrinsic defects or extrinsic defects related to common impurities that might be introduced involuntarily at the synthesis.  These defects are therefore the most commonly encountered ones when analyzing usual samples. However, the wide band gap of \cbn represents a solid platform to host optically active defects through deliberate doping. For this reason, there exists an active research in the field with most of works dedicated to rare earth which promote the appearance of very bright additional emission lines.\cite{nakayama_characterization_2005,Shishonok-rare-earth-05,SHISHONOK20071602,shishonok_photoluminescence_2010,zhigunov_photo-_2016}

\subsubsection{Nano-cathodoluminescence}\label{sec-nanoCL-exp}

The summary given here provides a global picture of the variety of optical 
active centers identified in \cbn. This richness is shared with other wide 
band-gap semiconductors such as diamond, SiC or GaN, whose optical response has 
been the object of a much larger number of studies. One of the most promising 
application for optical defects in bulk materials is their use as single-photon 
sources which are at the core of future quantum information technologies. During the last years, \hbn has appeared as a very promising new platform for single-photon emission (SPE). Bright, stable and room-temperature active SPEs have been identified in the visible\cite{tran_quantum_2016,tran_robust_2016,shotan_photoinduced_2016,
jungwirth_temperature_2016,chejanovsky_structural_2016,Martinez_efficient_2016,
exarhos_optical_2017} 
and UV\cite{bourrellier_nanometric_2014} energy region. However, the most studied emissions present a large energy distribution (1.7--2.3 eV) dependently on the defect environment. A reason for this behavior might be the layered structure of \hbn which allows very low energy, and thus frequent structural deformations such as plane glides, bends or dislocations. \cbn could be also a potential host for SPEs with the advantage over \hbn of presenting a smaller set of emission energies, as it was summarized in the previous paragraph. However, \cbn is a rather defective material and individual defects can hardly be isolated using conventional optical techniques. This problem has hindered up to now more in-depth studies and for instance lifetimes and quantum nature of the optical centers have not been investigated yet. Here we partially overcome this problem by employing nano-CL and we provide the first maps of the spatial localization of the defects and insights on their quantum nature.

\begin{figure*}[t!]
	\centering
	\includegraphics[width=0.93\textwidth]{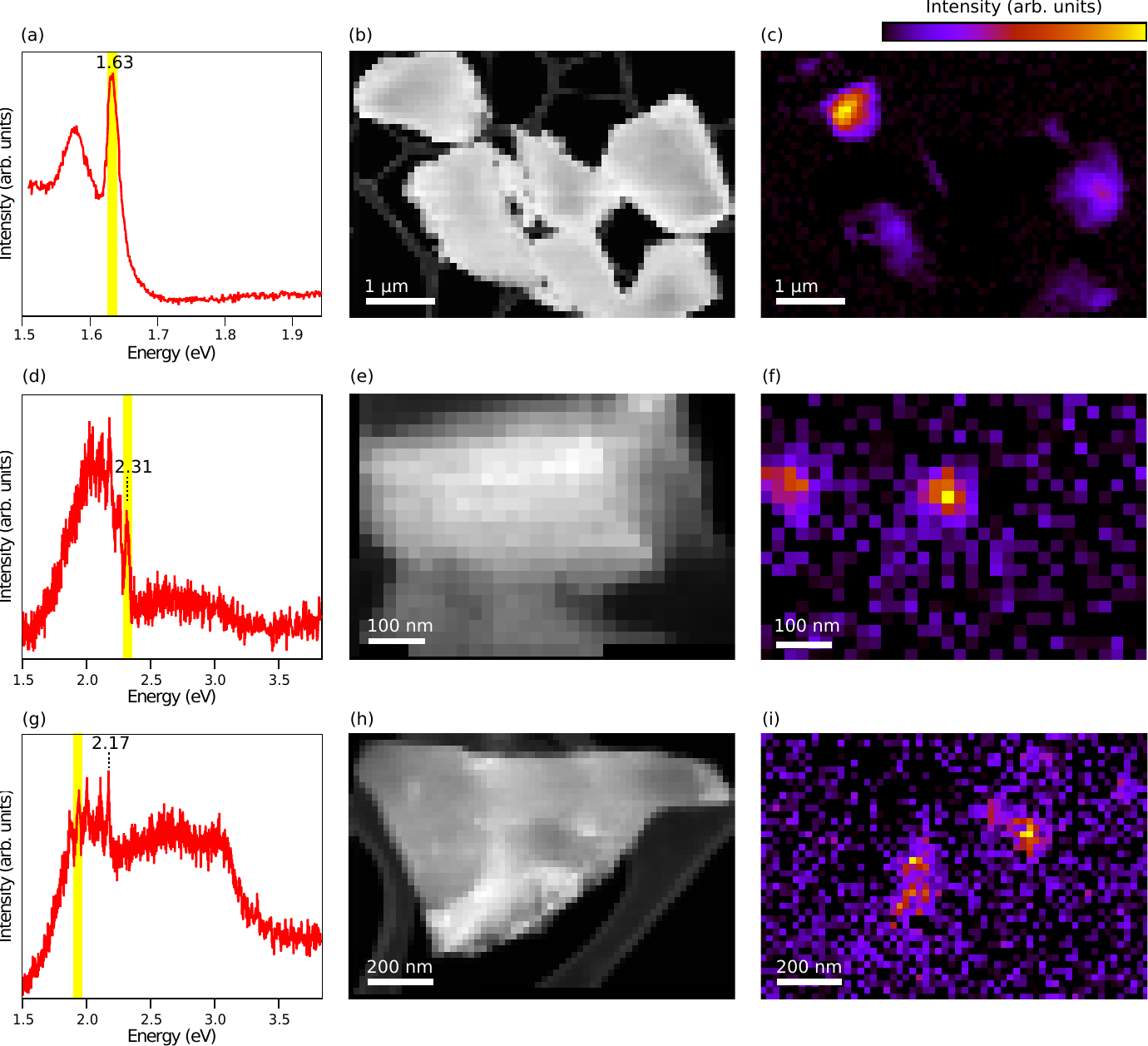}
	\caption{(a),(d),(g) Examples of typical spectra associated with defects in $c$--BN, (b),(e),(h) HAADF images of the particles from which they 
		have been extracted. (c),(f),(i) Corresponding CL emission intensity maps obtained by integrating the signal in the 10 meV window indicated by yellow regions in the spectra. For the spectra (a) and (g), the integration window correspond to zero-phonon lines. For the spectrum (d) a lower-energy phonon replica has been chosen in order to limit the overlap with the broad band centered at 2.5 eV.} 
	\label{All-emission}
\end{figure*}

By analyzing micro-particles of both high-quality and commercial \cbn crystals, several emission features have been clearly identified. In Fig. \ref{All-emission}(a,d,g) we  present some example spectra, in Fig. \ref{All-emission}(b,e,h) HAADF images from the particles from which they have been extracted and in  Fig. \ref{All-emission}(c,f,i) CL emission intensity maps. The zero-phonon line (ZPL) observed at 1.63 eV  (possibly a double peak) with a  56 meV phonon replica [Fig. \ref{All-emission}(a)] can be unequivocally identified as the GC--2 center. Two  ZPL at 2.31 eV [Fig. \ref{All-emission}(d)] and 2.17 eV [Fig. \ref{All-emission}(g)] can also be identified with multiple-phonon replica spaced by about 70 meV. Despite a minor energy shift, these defects can be reasonably associated to the RC--1 and RC--2 centers which have been observed in high-energy electron irradiated samples.\cite{zaitsev_cathodoluminescence_1986, shishonok_near-threshold_2002} In our experiments, these defects do not appear after deliberate prolongated irradiation with the microscope electron beam suggesting that the irradiation energy threshold should be above our 60 keV operating energy and that these defects might occur also as native defects. In Fig. \ref{All-emission}(d,g) two broad bands centered at about 2.5 eV and 3.12 eV are clearly visible. The first one resembles to the unidentified broad bands discussed in Ref. \citenum{shipilo_influence_1988}. The second one corresponds to the unresolved phononic structure of the PF--1 defect. 

Intensity maps have been derived by integrating the hyperspectral signal within energy windows of about 10 meV [Fig \ref{All-emission}(c,f,j)]. For the 1.63 eV and 2.31 eV emissions, integration windows were centered at the energy of the ZPL peaks while for the 2.17 eV emission we chose to center the window at a lower-energy phonon replica in order to limit the overlap with the broad bands. All emissions arise from highly spatially localized spots with the 2.17 eV and 2.31 eV centers having a FWHM below 100 nm, which is a value of the order of magnitude of excitation diffusion length in \hbn\cite{bourrellier_bright_2016} or diamond.\cite{Tizei2012,tizei_spatially_2013} Therefore, it can be reasonably proposed that, by analogy with the case of other wide band-gap semiconductors, these \cbn optically active centers could be related to point defects and ultimately act as single-photon sources. The relatively low quantum efficiency of these centers, whose spectra also often overlap, presents a strong limit to investigate this hypothesis using our nano-CL setup coupled with a HBT interferometer. High-energy emissions are more adapted to these studies since they present stronger luminescences and their spectra can be clearly isolated, which grants a higher signal-to-noise and a higher signal-to-background ratio. 

\begin{figure}[t]\centering
	\includegraphics[width=1.0\columnwidth]{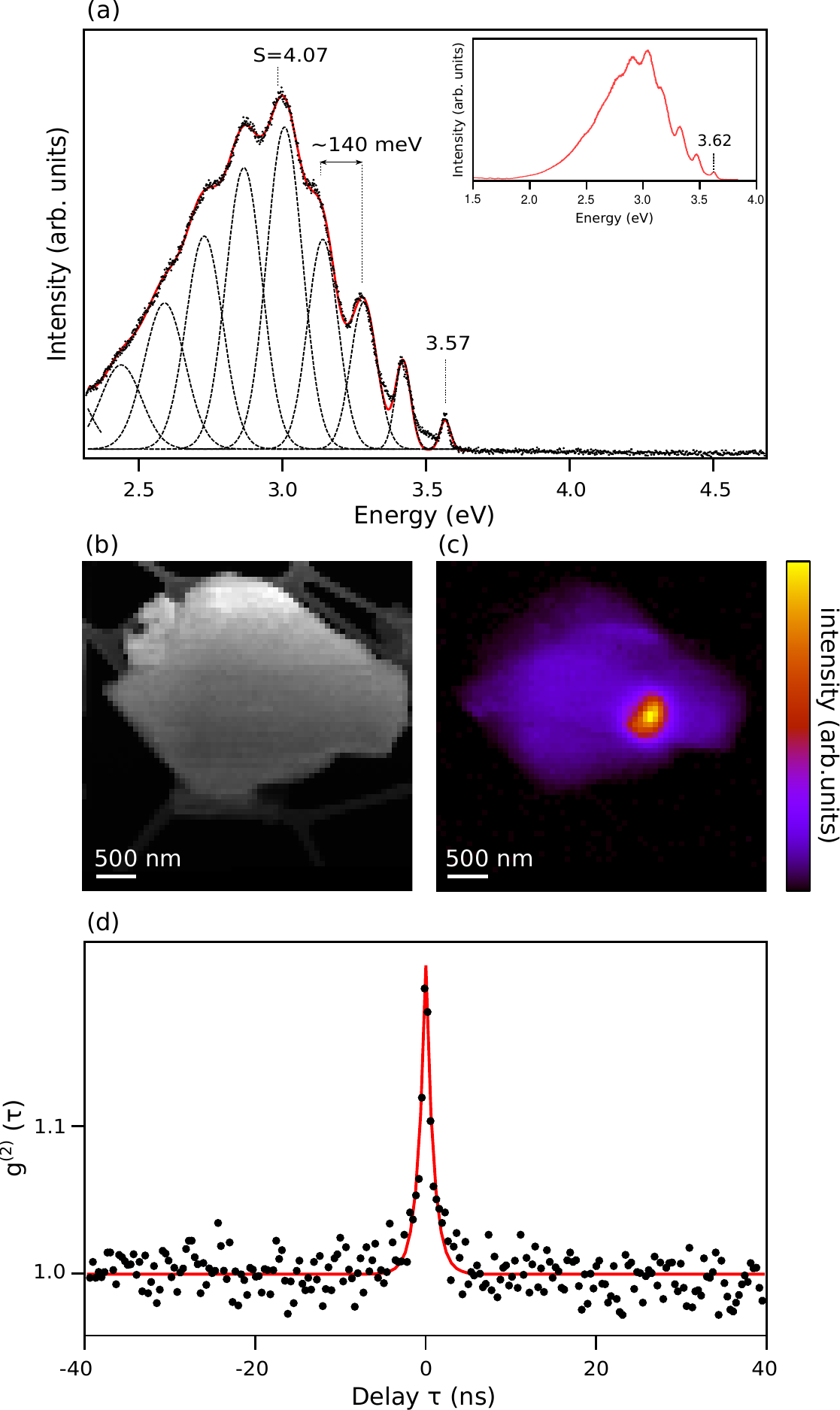}
	\caption{(a) The PF--1 zero-phonon line at 3.57 eV and associated phonon replica. Dashed curves are individual gaussian components used to fit the phonon replica, the red curve is the resulting global fit. In the insert we report a 3.62 eV emission which most likely arises from the same type of defect in a different local environment. (b) HAADF image of the $c$--BN 
	particle from which the spectrum has been extracted and (c) CL 	intensity map of the this emission. (d) Autocorrelation function derived from HBT 
	interferometry. An emission lifetime of $0.55\pm0.09$ ns is obtained from the fit of the bunching peak.} 
	\label{emissione-3-57eV-348nm}
\end{figure}

In Fig. \ref{emissione-3-57eV-348nm}(a) we present the emission spectrum of the 
PF--1 series with a ZPL clearly visible at 3.57 eV and six strong phonon replicas. The phonon structure has been fitted with a multi-Gaussian function set, giving an average line separation of 141 $\pm$ 6 meV. By comparison with the calculated phonon density of states,\cite{kern_ab_1999, karch_ab-initio_1997, bechstedt_dynamics_2000,bagci_ab_2006} this spacing corresponds to the LO phonon mode. It is also in agreement with the value measured in the EELS spectrum [Fig. \ref{full-spectrum}(a)].  Occasionally a slightly higher energy emission at 3.62 eV can be observed with phonon replicas having a little higher energy [see insert in Fig. \ref{emissione-3-57eV-348nm}(a)]. Both emissions probably arise from the same type of defect, and the energy shift may be due to a different local environment. Similar shifts were also observed in Ref. \citenum{shipilo_electron-vibrational_1990}. The deexcitation process can be characterized by  the Huang-Rhys factor $S$  which represents the average number of phonons involved in the emission. The energy $\hbar \omega_{\text{ZPL}} - S\hbar \omega_{\text{phon}}$, where $\hbar 
\omega_{\text{ZPL}}$ is the energy of the ZPL and $\hbar \omega_{\text{phon}}$ is the phonon energy, should indicate the maximum of the emission band. From our
fit we obtain a Huang-Rhys factor of 4.07 and hence a maximum of intensity 
expected at 2.99 eV in very good agreement with the experiments. A similar 
characterization had been reported in previous works
\cite{manfredotti_vibronic_2006,Shishonok-vibrations-2003} obtaining slighty higher values (4.9 and 5.26), but it should be kept in mind that the Huang-Rhys factor varies as a function of the local environment and therefore it is not an intrinsic parameter of the defect emission.

In Fig. \ref {emissione-3-57eV-348nm}(b) we present a HAADF image of the 
particle from which the spectrum in Fig. \ref {emissione-3-57eV-348nm}(a)  has 
been extracted. In Fig. \ref {emissione-3-57eV-348nm}(c) we report the emission 
intensity map obtained by integrating the signal in an energy window about 100 
meV wide and centered at the emission band maximum (3.00 eV). Whereas a faint signal is present all over the particle, once more it is possible to identify a localized emission spot about 500 nm wide indicating a limited number of emission centers with the possibility of isolated centers.

This emission has been further investigated by driving the CL signal to a unique home-made HBT interferometer. Fig. \ref{emissione-3-57eV-348nm}(d) reports a typical intensity autocorrelation function acquired from these centers. The curve shows an evident zero-delay bunching character. This behavior occurs in CL due to the synchronized emission of several defect centers excited by the same electron through the deexcitation of a bulk plasmon into few electron-hole pairs. This effect has been previously observed for point defects in diamond and \hbn\cite{meuret_photon_2015} and more recently in InGaN quantum wells.\cite{PhysRevB.96.035308} The bunching is 
characteristic of emissions by packets of photons: although anti-bunching is the irrefutable evidence of a SPE, in the particular case of CL, bunching is obtained from a set of independent but spatially close SPEs. Our experiments have not permitted yet to isolate a single defect associated with the 3.57 eV luminescence even in high-quality crystals. The relatively high spatial extension observed in the emission maps indicates the clustering of several defect centers within the typical plasmon-excitation volume. Finally, an analysis of the bunching decay provides the lifetime of the optical center.\cite{meuret_lifetime_2016} By fitting the experimental curve with exponential functions we estimate the lifetime of the 3.57 eV emission to be $1.4 \pm 0.1$ ns.

\begin{figure}[t]\centering
	\includegraphics[width=1.0\columnwidth]{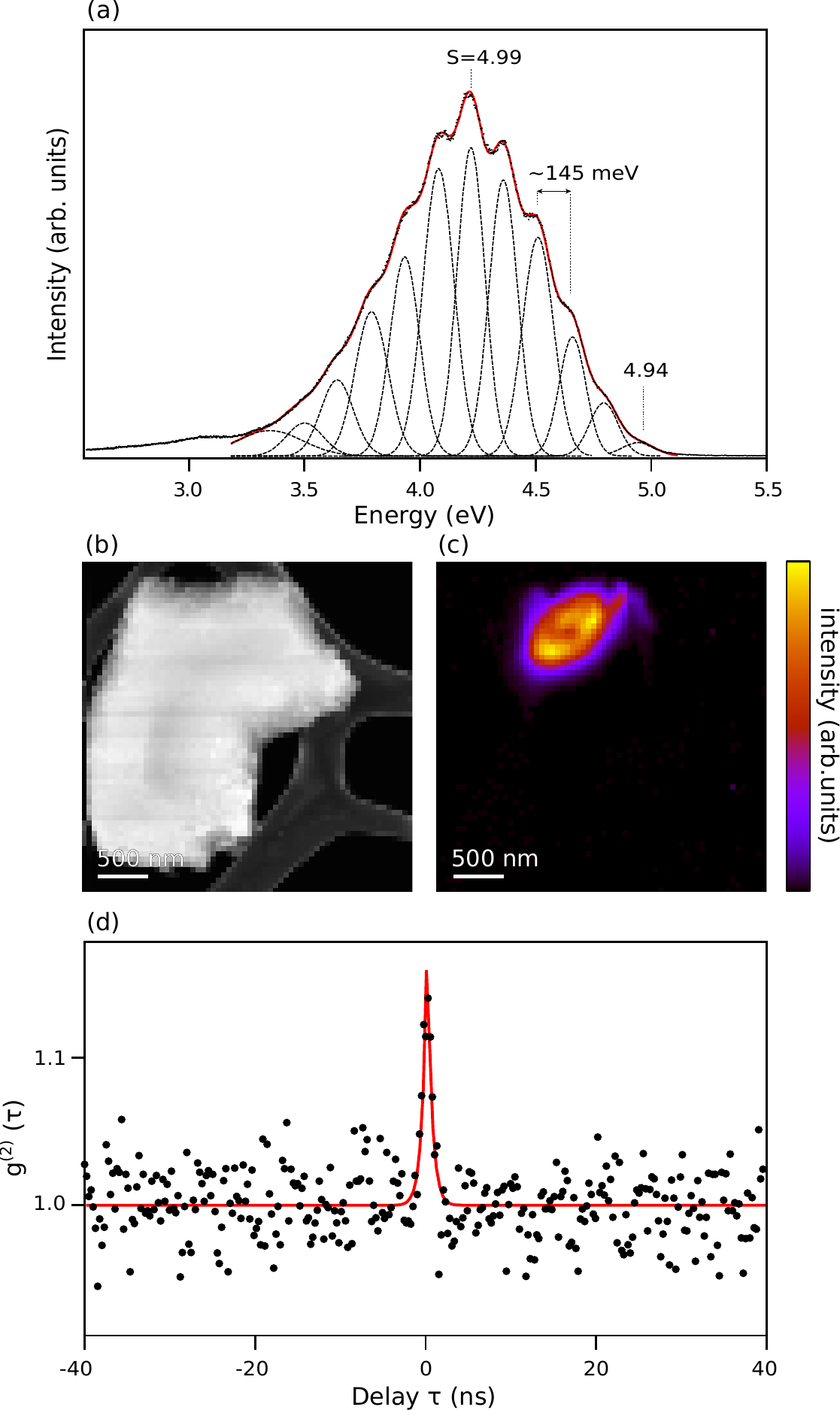}
	\caption{(a) The T defect zero-phonon line at 4.94 eV and associated phonon replica. Dashed curves are indivual gaussian components used to fit the phonon replica, the red curve is the resulting global fit. (b) HAADF image of the $c$--BN particle from which the spectrum has been extracted and (c) CL intensity map of this emission from which several separated spots can be distinguished. (d) Autocorrelation function derived from HBT interferometry. An emission lifetime of $1.4 \pm 0.1$ ns is obtained from the fit of the bunching peak.} 
	\label{emission-4.94eV}
\end{figure}

A similar analysis has been conducted for the T series having a ZPL at 4.94 eV (Fig. \ref{emission-4.94eV}). In the spectrum [Fig. \ref{emission-4.94eV}(a)], eight phonon replicas can be distinguished with an average line separation of $145\pm 5$ eV, once more compatible with LO phonons. The calculated Huang-Rhys factor is 4.99 from which one derives an energy for the maximum of the emission band of 4.22 eV, in good agreement with the experimental value. Fig. \ref{emission-4.94eV}(b,c) present a microscopy image of a whole \cbn particle and the associated intensity map for the 4.94 eV emission, derived by signal integration in a 100 meV energy window centered at the band maximum. At least three intensity maxima can be distinguished in the map, indicating a possible clustering of several defect centers. HBT experiments have been performed and a typical autocorrelation function is presented in Fig. \ref{emission-4.94eV}(d). Once more we obtain a bunching function indicating the possible presence of multiple close SPEs. From the exponential fit we estimate the lifetime of the center as $0.55\pm0.09$ ns.

Nano-CL experiments have also been conducted on Tb and Eu earth doped \cbn crystals (not presented here). We observe a strong catholuminescence related to the rare earth atoms analogous to previously published results.\cite{nakayama_characterization_2005} However related nano-CL maps  show a luminescence delocalized all over the particles indicating a high doping level which does not allow to isolate individual emitters.

\section{Conclusions}

In this work we investigated the optical properties of \cbn tackling both the problem of the  optical  gap and the luminescence of intragap defects. By using EELS  with a monochromated electron beam on high-quality \cbn crystals we demonstrated unequivocally that the optical gap slightly exceeds 10 eV. A further theoretical analysis confirms this result permitting to interpret the nature of the spectral features observed. The main peak corresponds to a plasmon excitation while the onset of the optical response is identified as a Wannier-type exciton. Excitonic effects modulate the intensity of the spectra and they can be associated to several features observed in experiments. Previous estimation of the optical gap, corresponding to much lower values, should be understood as misinterpretations of optically active defects, possibly related to inclusions of hexagonal phases. By high-temperature annealing a \cbn powder we have deliberately promoted phase transitions.  In very local domains of these treated samples, we observed in correlated EELS and CL experiments the appearance of spectroscopic signatures characteristic of the hexagonal phase. 

A high number of  optically active centers are known in \cbn but the literature is rather scattered and hard to follow. Here we summarized the main known results with the aim to provide an unified picture of defect luminescence in \cbn. Using our unique nano-CL experimental setup integrated within a STEM, we have detected a number of these centers in the visible and UV spectral range. Due to the large band gap of \cbn, higher-energy optically active defects might also be present. These transitions can in principle be excited 
by the microscope's high energy electron beam but they cannot be observed by
our CL system which has an upper detection limit of about 6.2 eV.
Finally, we have investigated by nano-CL the spatial localization of several emission centers showing small spots, compatible with a limited number of optically active centers. For the brightest defect types, lying in the UV spectral range, we have performed Hanbury Brown and Twiss interferometry experiments in order to identify the possible presence of single-photon sources.
The bunching effect observed is compatible with a very limited set of SPEs synchronously excited but no anti-bunching has been detected yet. We hope that these promising results will motivate additional investigations which might succeed in isolating individual centers. Moreover, SPEs might be associated with emissions in the visible spectral range which present also a high spatial localization. In this work, we have not tested these centers by HBT interferometry due to their relatively low CL-luminescence and a spectral overlap with the tail of higher energy emissions inevitably excited by CL with fast electrons. Photoluminescence in a confocal microscope might an adapted technique whereas a too high density of defects in currently available samples might still hinder these studies.

\begin{acknowledgments}
 AT, MK, TZ and AZ acknowledge support from the Agence Nationale de la Recherche (ANR), program of future investment TEMPOS-CHROMATEM (No. ANR-10-EQPX-50). The  work has also received funding from the European Union in Seventh Framework  Programme (No. FP7/2007 -2013) under Grant Agreement No. n312483 (ESTEEM2), from a Marie Curie FP7 Integration Grant within the 7th European Union Framework Programme and from the European  Research Council (ERC Grant Agreement n. 320971). 	Computational time was granted by GENCI (Project No. 544).
\end{acknowledgments}
\vfill

\bibliography{biblio}

\end{document}